\documentclass[twoside,twocolumn,9pt]{article}
\usepackage{extsizes}
\usepackage[super,sort&compress,comma]{natbib} 
\usepackage[version=3]{mhchem}
\usepackage[left=1.5cm, right=1.5cm, top=1.785cm, bottom=2.0cm]{geometry}
\usepackage{balance}
\usepackage{times,mathptmx}
\usepackage{sectsty}
\usepackage{graphicx} 
\usepackage{lastpage}
\usepackage[format=plain,justification=justified,singlelinecheck=false,font={stretch=1.125,small,sf},labelfont=bf,labelsep=space]{caption}
\usepackage{float}
\usepackage{fancyhdr}
\usepackage{fnpos}
\usepackage[english]{babel}
\usepackage{array}
\usepackage{droidsans}
\usepackage{charter}
\usepackage[T1]{fontenc}
\usepackage[usenames,dvipsnames]{xcolor}
\usepackage{setspace}
\usepackage[compact]{titlesec}

\usepackage{amssymb}
\usepackage{amsmath}

\usepackage{epstopdf}

\definecolor{cream}{RGB}{222,217,201}

\begin{document}

\thispagestyle{plain}


\makeFNbottom
\makeatletter
\renewcommand\LARGE{\@setfontsize\LARGE{15pt}{17}}
\renewcommand\Large{\@setfontsize\Large{12pt}{14}}
\renewcommand\large{\@setfontsize\large{10pt}{12}}
\renewcommand\footnotesize{\@setfontsize\footnotesize{7pt}{10}}
\makeatother

\renewcommand{\thefootnote}{\fnsymbol{footnote}}
\renewcommand\footnoterule{\vspace*{1pt}%
\color{cream}\hrule width 3.5in height 0.4pt \color{black}\vspace*{5pt}} 
\setcounter{secnumdepth}{5}

\makeatletter 
\renewcommand\@biblabel[1]{#1}            
\renewcommand\@makefntext[1]%
{\noindent\makebox[0pt][r]{\@thefnmark\,}#1}
\makeatother 
\renewcommand{\figurename}{\small{Fig.}~}
\sectionfont{\sffamily\Large}
\subsectionfont{\normalsize}
\subsubsectionfont{\bf}
\setstretch{1.125} 
\setlength{\skip\footins}{0.8cm}
\setlength{\footnotesep}{0.25cm}
\setlength{\jot}{10pt}
\titlespacing*{\section}{0pt}{4pt}{4pt}
\titlespacing*{\subsection}{0pt}{15pt}{1pt}

\setlength{\arrayrulewidth}{1pt}
\setlength{\columnsep}{6.5mm}
\setlength\bibsep{1pt}

\makeatletter 
\newlength{\figrulesep} 
\setlength{\figrulesep}{0.5\textfloatsep} 

\newcommand{\topfigrule}{\vspace*{-1pt}%
\noindent{\color{cream}\rule[-\figrulesep]{\columnwidth}{1.5pt}} }

\newcommand{\botfigrule}{\vspace*{-2pt}%
\noindent{\color{cream}\rule[\figrulesep]{\columnwidth}{1.5pt}} }

\newcommand{\dblfigrule}{\vspace*{-1pt}%
\noindent{\color{cream}\rule[-\figrulesep]{\textwidth}{1.5pt}} }

\makeatother

\renewcommand{\vec}[1]{\boldsymbol{\mathbf{#1}}}
\newtheorem{theorem}{Theorem}

\twocolumn[
  \begin{@twocolumnfalse}
\sffamily
\begin{tabular}{m{2.5cm} p{13.5cm} }

 & \noindent\LARGE{
 \textbf{
Maximum likelihood estimations of force and mobility from short single Brownian trajectories
}
} 
\\
\vspace{0.3cm} & \vspace{0.3cm} 
\\

& \noindent\large{
Raphael Sarfati,\textit{$^{a}$} 
Jerzy B\l{}awzdziewicz,\textit{$^{b}$} 
and 
Eric R. Dufresne$^{\ast}$\textit{$^{c}$}} 
\\

& \noindent\normalsize{
We describe a  method to extract force and diffusion parameters from single trajectories of Brownian particles based on the principle of maximum likelihood.
The analysis is well-suited for out-of-equilibrium trajectories, even when a limited amount of data is available and the dynamical parameters vary spatially.
We substantiate this method with experimental and simulated data, and discuss its practical implementation, strengths, and limitations.
}

\end{tabular}
\end{@twocolumnfalse} \vspace{0.6cm}
]

\renewcommand*\rmdefault{bch}\normalfont\upshape
\rmfamily
\section*{}
\vspace{-1cm}


\footnotetext{\textit{$^{a}$~Department of Applied Physics, Yale University, New Haven, CT 06520, USA. }}
\footnotetext{\textit{$^{b}$~Department of Mechanical Engineering, Texas Tech University, Lubbock, TX 79409, USA. }}
\footnotetext{\textit{$^{c}$~Department of Materials, ETH Z\"{u}rich, 8092 Z\"{u}rich, Switzerland. 
E-mail: ericd@ethz.ch}}




\section{Introduction}

Brownian particles are ubiquitous in soft matter and biological sciences, from colloidal particles to fluorescently-tagged molecules.
The trajectories of these particles contain precious information about their structure and interactions.
For example, particle sizes are routinely quantified  by measuring diffusion coefficients of a dilute suspension in a well-characterized fluid\cite{clark1970,DLS}.
Alternatively, when the particles are well-characterized, the trajectory of a Brownian particle can probe the solvent's rheological properties\cite{mason}.
Analysis of Brownian trajectories can also reveal the conservative and dissipative forces acting on particles due to external fields\cite{roichman2008} or interactions with other particles\cite{crocker1994}.

Since Brownian trajectories are stochastic, their analysis is necessarily statistical.
The physical theory describing the statistics of Brownian particles is well-established\cite{perrin}.
When particles fluctuate near an equilibrium position, conservative forces acting on the particle are readily extracted using Boltzmann statistics\cite{prieve1990}.
In the absence of external forces, the dissipative forces acting on the particle can be calculated from its diffusion coefficient using the Stokes-Einstein relation.
More generally, the Smoluchowski equation can describe the trajectory of Brownian particles when conservative force and diffusion coefficient vary over space\cite{chandrasekhar1943}.
Existing methods based on the Smoluchowski equation require many trajectories so that the distribution of step sizes at each location can be robustly sampled.
This approach has been successfully implemented for micron-sized colloidal particles, where they are typically trapped and released repeatedly using optical traps\cite{crocker1994,sainis2007,merrill2010,evans2016}.
However, most Brownian particles cannot be manipulated with optical traps:  
they are either too small, or do not have appropriate optical properties.
In these cases, one requires a method to determine the forces with much less data.
We have recently introduced a new method to analyze single Brownian trajectories, based on the general principle of maximum likelihood\cite{sarfati}, capable of extracting dynamical parameters from individual trajectories.
The idea is to perform a global fit of a trajectory, instead of a local one.
The metric is probabilistic: the method searches for the most likely force and diffusion profiles to have resulted in the observed trajectory.

In this paper, we illustrate maximum likelihood analysis (MLA) applied to Brownian trajectories. 
First, we provide a simple presentation of the principles of the analysis, based on probabilistic considerations.
Second, we demonstrate the method based on simulations and experimental data, including the analysis of strong interactions between pairs of paramagnetic particles.
Finally, we discuss the practical implementation of the method.

%
%

\section{Theoretical Background}

In this section, we review the solution to the Smoluchowski equation and the principle of maximum likelihood. 
We illustrate how to combine these two concepts on a simple example.

Let us consider the one-dimensional position $x$ of a particle over time, and call $F$ the applied force and $D$ the diffusion coefficient.
In general, $F$ and $D$ may depend on $x$.
However, on sufficiently short time intervals $\Delta t$ the particle samples a region where $F$ and $D$ are uniform.
In this case, the solution of the Smoluchowski equation states that the displacements $\delta x = x(t+\Delta t) - x(t)$ over $\Delta t$ follow a Gaussian distribution of mean $\mu_{\Delta t}$ and variance $\sigma^2_{\Delta t}$ given by
\begin{equation}
\mu_{\Delta t} = \overline{v} \Delta t,
\label{mu}
\end{equation}
\begin{equation}
\sigma^2_{\Delta t} = 2D \Delta t, 
\label{sigma2}
\end{equation}
where $\overline{v}$ is the drift velocity:
\begin{equation}
\overline{v} = bF + \nabla D.
\end{equation}
Here, $b$ is the mobility and is related to $D$ by the Stokes-Einstein relation:
\begin{equation}
D = b k_B T,
\end{equation}
with $k_B$ the Boltzmann constant and $T$ the absolute temperature.

\begin{figure}[t]
\includegraphics[]{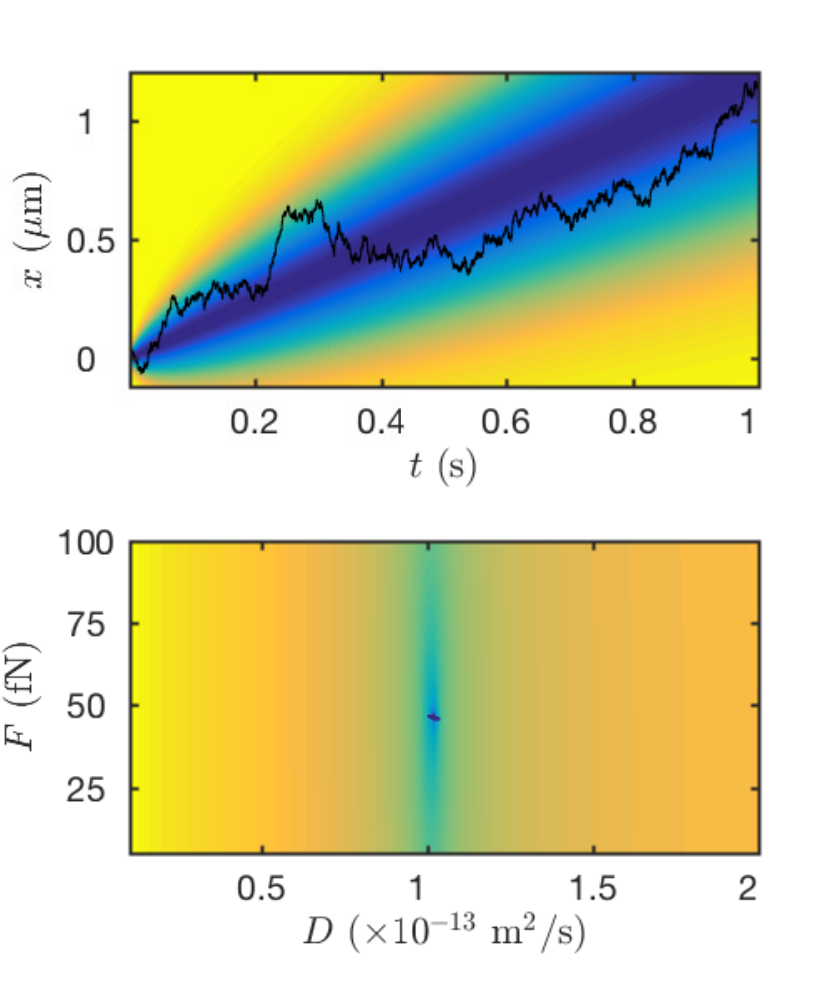}
\centering
\setlength{\unitlength}{1cm}
\put(-7,8.7)
{\large \text{(a)}}
\put(-7,4)
{\large \text{(b)}}
\caption{\label{rnd} 
Brownian basics.
(a) 
Probability density (normalized at each time point for visibility) of observing a room temperature Brownian particle with $D = 10^{-13}$~m$^2$/s and $F = 50$~fN at position $x$ at time $t$, starting from $x(0)=0$, and simulated random trajectory (black line).
Darker colors indicate higher probability.
(b)
Log-likelihood landscape in the $(D,F)$ plane associated with the trajectory in (a).
For better visibility, the log-likelihood is renormalized by the maximum value on the grid.
Darker colors indicate higher log-likelihood values.
}
\end{figure}

The time evolution of the probability distribution for a constant force and diffusion coefficient is illustrated in Fig.~\ref{rnd}a.
From an experimental trajectory $X = \lbrace x_1, \dotsc , x_{N+1} \rbrace$ with time interval $\Delta t$, $F$ and $D$ can be directly calculated from the mean and variance of the displacements, $\delta x_i = x_{i+1}-x_i$, according to eqns (\ref{mu},\ref{sigma2}).

In contrast, the principle of MLA is to identify the set of dynamical parameters, $\pmb{\alpha}$, that are most likely to describe an observed set of displacements, $\delta~=~\lbrace \delta x_i \rbrace_{i=1\dotsc N}$. 
The likelihood $\mathcal{P}(\pmb{\alpha} \, \vert \, \delta)$ that $\pmb{\alpha}$ describes the observed displacements $\delta$ can be expanded as
\begin{equation}
\mathcal{P}(\pmb{\alpha} \, \vert \, \delta) 
= \prod_{i=1}^{N} p(\pmb{\alpha} \, \vert \, \delta x_i).
\label{product}
\end{equation}
Here, $p(\pmb{\alpha} \, \vert \, \delta x_i)$ is the probability that $\pmb{\alpha}$ underlies the particle dynamics given a single observed  displacement $\delta x_i$.
(Eqn~(\ref{product}) assumes the displacements are independent. 
This is true in the non-inertial regime, where the displacements are measured over a time $\Delta t \gg m/\gamma$, with $m$ the particle mass and $\gamma$ its drag coefficient. 
The inertial relaxation time scales with the square of the particle diameter.  
For micrometric particles in water, $m/\gamma \sim 10^{-6}$~s.)
Smoluchowski's theory provides the probability of observing a specific displacement given a set of parameters, $p(\delta x_i \, \vert \, \pmb{\alpha})$.
Using Bayes' theorem, we can determine  $p(\pmb{\alpha} \, \vert \, \delta x_i)$:
\begin{equation}
p(\pmb{\alpha} \, \vert \, \delta x_i) 
= \frac{p(\delta x_i)}{p(\pmb{\alpha})} p(\delta x_i \, \vert \, \pmb{\alpha}).
\end{equation}
Assuming  the priors $p(\pmb{\alpha})$ and $p(\delta x_i)$ to be uniform on some reasonable intervals, the likelihood that a set of parameters $\pmb{\alpha}$ describes an observed set of displacements $\delta$ reads
\begin{equation}
\mathcal{P}(\pmb{\alpha} \, \vert \, \delta) 
= \omega \prod_{i=1}^{N} p(\delta x_i \, \vert \, \pmb{\alpha}),
\end{equation}
where $\omega$ is a constant.

For numerical stability, it is more convenient to work with log-likelihood functions. 
For an observed trajectory $X$ of corresponding set of displacements $\delta$, we define the log-likelihood function of argument $\pmb{\alpha}$ as
\begin{equation}
L_\delta (\pmb{\alpha}) 
= \ln \left( \omega^{-1} \mathcal{P}(\pmb{\alpha} \, \vert \, \delta ) \right)
= \sum_{i=1}^{N} \ln p(\delta x_i \, \vert \, \pmb{\alpha})
\end{equation}
Is it important to note that $L_\delta (\pmb{\alpha})$ depends on $\delta$, that is, on the trajectory considered.

The determination of $\pmb{\alpha} = (F,D)$ from a single Brownian trajectory is illustrated in Fig.~\ref{rnd}.
A simulated trajectory for a room-temperature Brownian particle with $D = 10^{-13}$~m$^2$/s and $F = 50$~fN is super-imposed over the probability distribution as the black curve in Fig.~\ref{rnd}a.
The log-likelihood landscape associated to this trajectory in the $\pmb{\alpha} = (F,D)$ phase space is plotted in Fig.~\ref{rnd}b.
The log-likelihood function is maximized at the appropriate values of $F$ and $D$, as visible in Fig.~\ref{rnd}b and mathematically supported in the ESI$^\dag$.  

The utility and limitations of this approach are made apparent in the following sections.

\section{Illustrative Examples}

\subsection{Experimental trajectory of a single particle with constant $F$ and $D$}

\begin{figure}[t]
\vspace{-1cm}
\includegraphics[width = 0.97\linewidth]{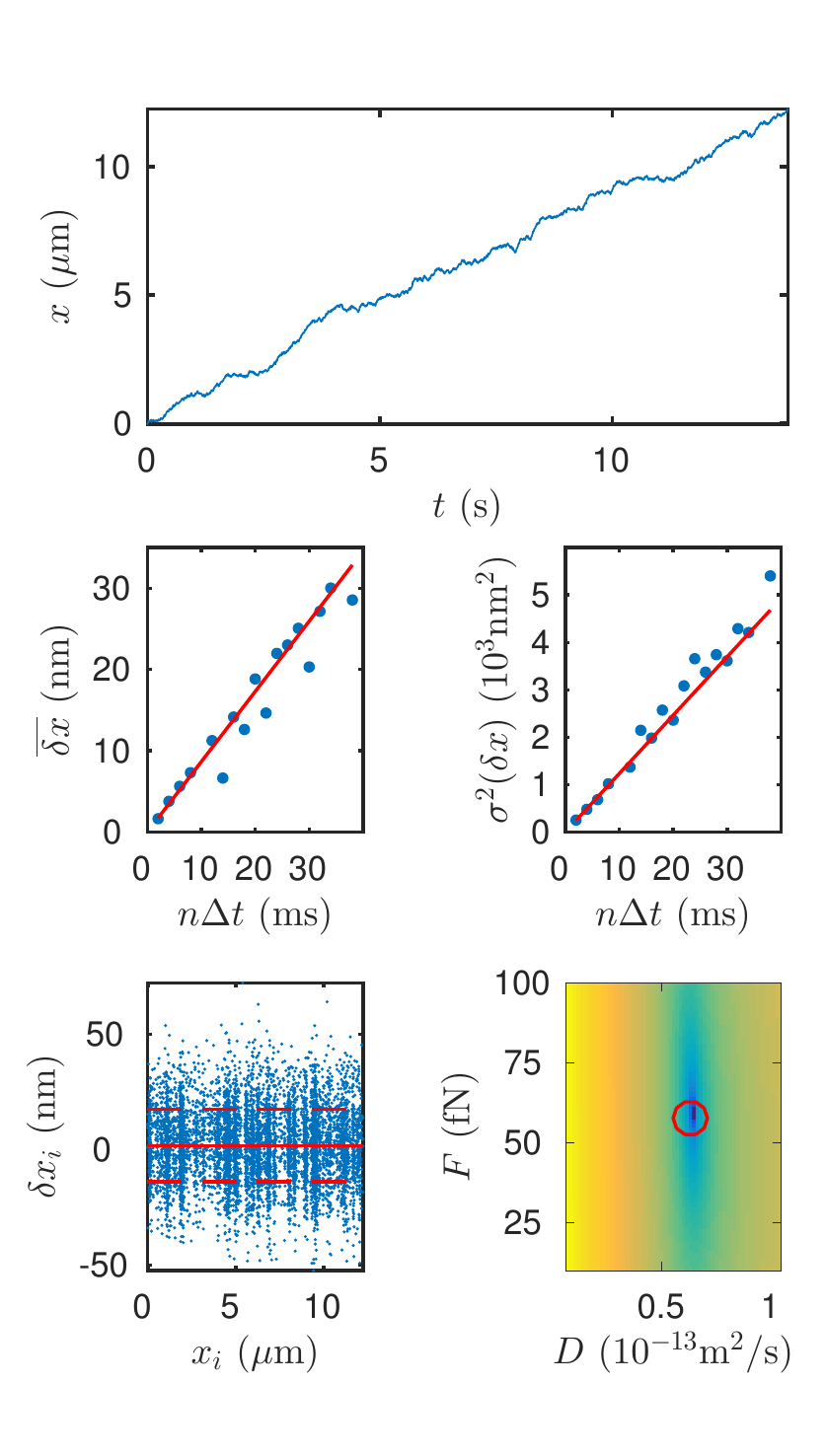}
\centering
\setlength{\unitlength}{1cm}
\put(-3.3,11.2)
{\includegraphics[height=1.3cm]{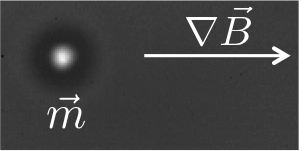}}
\vspace{-1cm}
\put(-7,13.7)
{\large \text{(a)}}
\put(-7,9.1)
{\large \text{(b)}}
\put(-2.6,9.1)
{\large \text{(c)}}
\put(-7,4.6)
{\large \text{(d)}}
\put(-2.6,4.6)
{\large \text{(e)}}
\put(-6.5,7.6)
{\Large \text{\scalebox{3}{$\lrcorner$} $\, \overline{v}$}}
\put(-1.9,7.6)
{\Large \text{\scalebox{3}{$\lrcorner$} $ 2D$}}
\caption{\label{fig:1pB} 
Comparison of MLA and conventional analysis with an experimental trajectory (a) of a paramagnetic particle in a constant magnetic field gradient (Inset). 
Mean (b) and variance (c) of the displacements $\delta x$ at different lag times $n\Delta t$ (blue dots), and corresponding linear fits (red lines).
The slopes are equal to $\overline{v}$ and $2D$, respectively.
(d) Frame-to-frame displacements $\delta x_i$ as a function of position $x_i$. 
Red line indicates MLA fit to the mean (1.78~nm), and red dashed lines MLA fit to the standard deviation ($\pm$~15.8~nm).
(e) Log-likelihood landscape (renormalized) of the trajectory in (a).
Red circle indicates results from statistical analysis.
}
\end{figure}
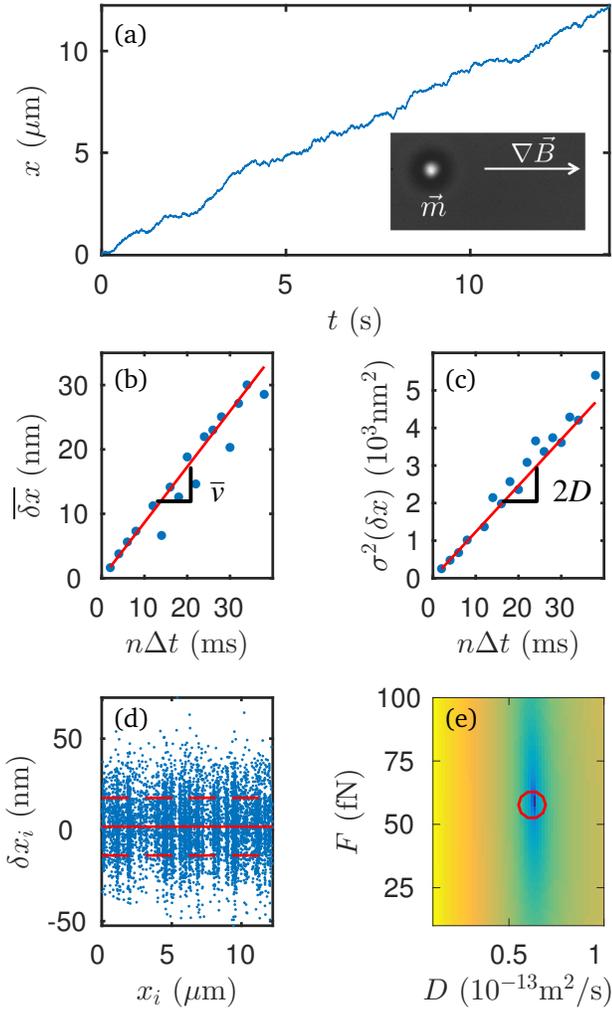

We compare measurements of the force and diffusion coefficent of a single Brownian particle using two different methods: 1) direct calculation of the time-dependent mean and variance of the trajectory \cite{sainis2007}, and 2) MLA. 
Paramagnetic spheres (2.8~$\mu$m diameter) are suspended in water and sediment against a glass coverslip. 
A permanent magnet is positioned next to the sample to create a locally uniform gradient of the magnetic field $B$ along the $x$-direction.
The field gradient drives the paramagnetic particle with constant external force.
Further experimental details are provided in the ESI.
A representative trajectory, $x(t)$, of the particle is shown in Fig.~\ref{fig:1pB}a.
Here, the frame-to-frame time interval is $\Delta t = 2$~ms, and the exposure time is $\tau_{ex} = 0.5$~ms.

From this trajectory, we calculate $\overline{v}$ (Fig.~\ref{fig:1pB}b) and $D$ (Fig.~\ref{fig:1pB}c) from (respectively) the mean and variance of the displacements at different lag times $\delta x = x_{i+n}-x_{i} = x(t_i+n\Delta t) - x(t_i)$.
We obtain the following estimations: $F = 58 \pm 3 $~fN and $D = 6.16 \pm 0.14 \times 10^{-14}$~m$^2$/s.

Alternatively, we can apply MLA to the distribution of the frame-to-frame displacements $\lbrace \delta x_i = x_{i+1}-x_i \rbrace$, using parametrization $\pmb{\alpha} = (F,D)$.
The results for the MLA fits of the mean and standard deviation are presented in Fig.~\ref{fig:1pB}d.
The log-likelihood landscape for this trajectory in the $(F,D)$ plane is represented in Fig.~\ref{fig:1pB}e. 
It shows a maximum at $(F,D) = (59 \pm 2$~fN, $6.28 \pm 0.15 \times 10^{-14}$~m$^2$/s), which corresponds to the values obtained from the conventional statistical analysis.
This demonstrates that MLA is reliable in this simple case of constant $F$ and~$D$.


\subsection{Simulated trajectories of pair interactions \label{sec:pairsim}}

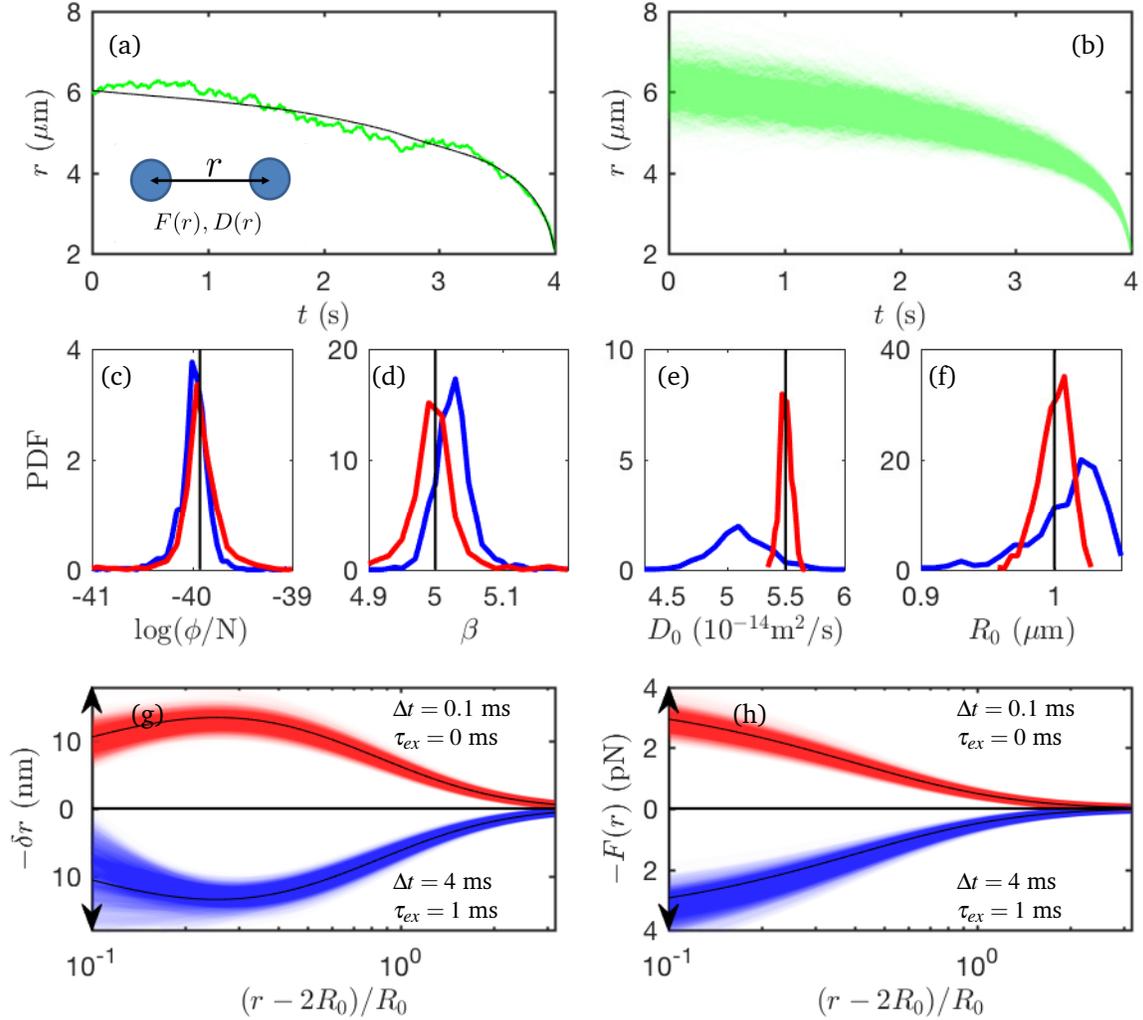
\begin{figure*}[t]
\vspace{-0.5cm}
\includegraphics[]{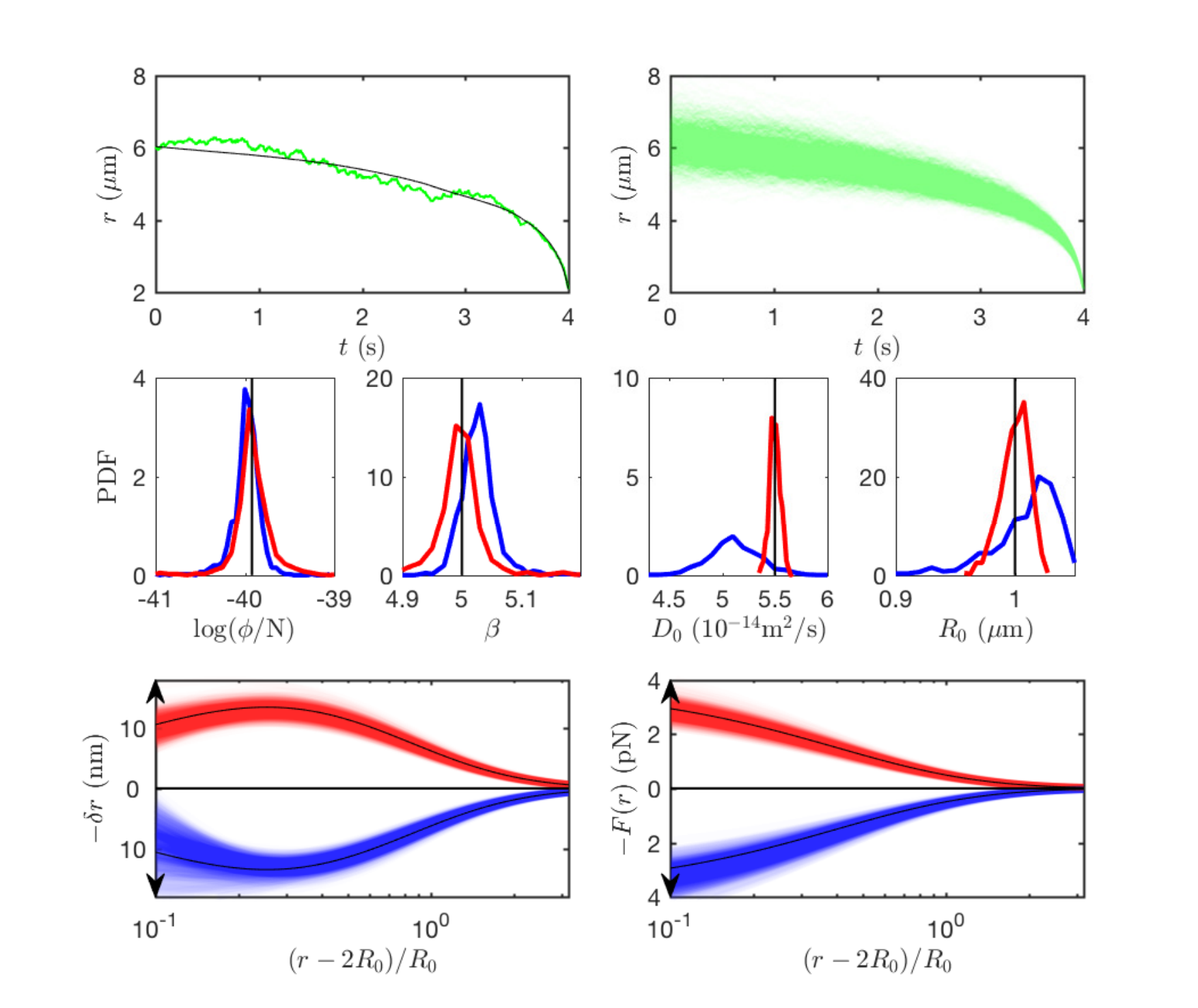}
\centering
\setlength{\unitlength}{1cm}
\put(-15.2,10.8)
{\includegraphics[height=1.3cm]{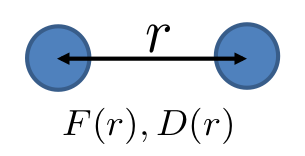}}
\put(-11.5,2)
{\parbox[]{2cm}{$\Delta t = 4$~ms \\ $\tau_{ex} = 1$~ms}}
\put(-11.5,4.3)
{\parbox[]{2cm}{$\Delta t = 0.1$~ms \\ $\tau_{ex} = 0$~ms}}
\put(-4,2)
{\parbox[]{2cm}{$\Delta t = 4$~ms \\ $\tau_{ex} = 1$~ms}}
\put(-4,4.3)
{\parbox[]{2cm}{$\Delta t = 0.1$~ms \\ $\tau_{ex} = 0$~ms}}
\put(-15.3,13.3)
{\large \text{(a)}}
\put(-2.5,13.3)
{\large \text{(b)}}
\put(-15.4,8.9)
{\large \text{(c)}}
\put(-11.8,8.9)
{\large \text{(d)}}
\put(-8,8.9)
{\large \text{(e)}}
\put(-4.4,8.9)
{\large \text{(f)}}
\put(-15,4.4)
{\large \text{(g)}}
\put(-7,4.4)
{\large \text{(h)}}
\caption{\label{fig:simul} 
Testing MLA for spatially varying force and diffusion coefficients with simulated trajectories.
(a) Sample trajectory (green line), and corresponding MLA fit (black line) from a simulation of colloidal spheres interacting via (Inset) force $F(r)$ and diffusion $D(r)$ as described in eqn~(\ref{F_pl},\ref{D_isotr}).
(b) All simulated trajectories. Darker shades signify  a higher density of curves.
(c-f) Probability density function (PDF) of the fit parameters obtained from MLA. 
Red curves correspond to trajectories with $\Delta t = 0.1$~ms and $\tau_{ex} = 0$~ms. 
Blue curves correspond to trajectories with $\Delta t = 4$~ms and $\tau_{ex} = 1$~ms. 
(g-h) Fitted displacement (g) and force (h) profiles obtained from MLA for all trajectories.
Black curves correspond to input profiles.
}
\end{figure*}

We focus now on the more complex case of pair interactions.
We assume that both the interparticle force, $F$, and the relative diffusion coefficient, $D$, depend on the center-to-center distance $r$ (Fig.~\ref{fig:simul}a-inset).
We perform simulations corresponding to pairs of micron-sized particles interacting through capillary interaction with a rough contact-line.
The  force profile is given by
\begin{equation}
F(r) = -\phi (x/a)^{-\beta},
\label{F_pl}
\end{equation}
with $\beta = 5$, following well-known theories \cite{stamou}, and $a~= 1$~m a scaling factor (necessary for dimensional consistency).
We choose a diffusion coefficient dependence corresponding to two particles in an isotropic fluid \cite{biancaniello}
\begin{equation}
D(r) = D_0 \times \dfrac{12(r/R_0-2)^2 + 8(r/R_0-2)}{6(r/R_0-2)^2 + 13(r/R_0-2) + 2},
\label{D_isotr}
\end{equation}
which works well for a capillary interaction when the two fluids have the same viscosity.
Here, $D_0$ is the 
one-particle diffusion coefficient at infinite separation, and $R_0$ is the hydrodynamic radius of the particle.

We perform $N_s = 1000$ simulations with timestep of $\Delta t_s = 0.1$~ms.
These trajectories depend on 4 parameters $\pmb{\alpha} = \lbrace \phi, \beta, D_0, R_0 \rbrace$. 
The input values are $\lbrace 1.2 \times 10^{-40}$~N$, 5, 5.5 \times 10^{-14}$~m$^2$/s, 1$\times 10^{-6}$~m $\rbrace$.
A single simulated trajectory is shown in Fig.~\ref{fig:simul}a, along with its corresponding MLA fit.
It is important to note that due to the stochastic nature of Brownian motion, each trajectory is different.
We plot a superposition of all trajectories in Fig.~\ref{fig:simul}b to show this variability.

We estimate the underlying physical parameters for each trajectory using MLA.
Details on the numerical optimization are given in the ESI.
As shown in Fig.~\ref{fig:simul}c-f, the MLA estimates (red histograms) agree very well with the input parameters (vertical black line).
The spatial dependence of the particle displacements and the force are also accurately captured by MLA, as shown in Fig.~\ref{fig:simul}g-h.
The red bands are a super-position of all the fitted profiles, and the black lines show the input to the simulation.  
The estimated profiles match very well the input profiles, with most of the deviations not exceeding about 10\% of the actual profiles.
We discuss the effect of trajectory blurring, present in video microscopy experiments, on the accuracy of MLA estimates in Section \ref{sec:data}.

These results show that MLA gives a very good estimation of the actual force profile and dynamic parameters for simulated trajectories.

\subsection{Experimental trajectory of magnetic dipole-dipole pair interactions}

We now investigate the reliability of MLA on experimental trajectories of isolated pairs of paramagnetic spheres in a magnetic field~$B$.
The magnetic field induces a magnetic dipole in both spheres, causing them to attract each other.
Neglecting the mutual induced dipoles effect occurring at short distances, the interaction is well described by a power-law force (eqn~(\ref{F_pl})) with exponent\cite{biswal} $\beta = 4$.
Because the beads are heavy and sedimented, their hydrodynamic coupling should also include a contribution from the bottom surface of the observation chamber. 
For the diffusion coefficient of the separation, we use a three-parameter functional form:
\begin{equation}
D(r) = \left( D_r^{-1}(r,D_0,R_0) + (D_h^{-1} - D_0^{-1})/2 \right)^{-1}
\label{D_wall}
\end{equation}
where $D_r(r,D_0,R_0)$ is given by eqn~(\ref{D_isotr}), $D_0$ and $R_0$ are defined as before, and $D_h$ is the diffusion coefficient of an isolated, sedimented bead at a small distance $h$ from the bottom surface \cite{dufresne2000}.
As we will see later, this simple three-parameter form matches the exact numerical solution of the Stokes equations \cite{jerzy2005} well in the range of moderate interparticle distances considered in this paper.

\begin{figure*}[]
\vspace{-1cm}
\includegraphics[width = \linewidth]{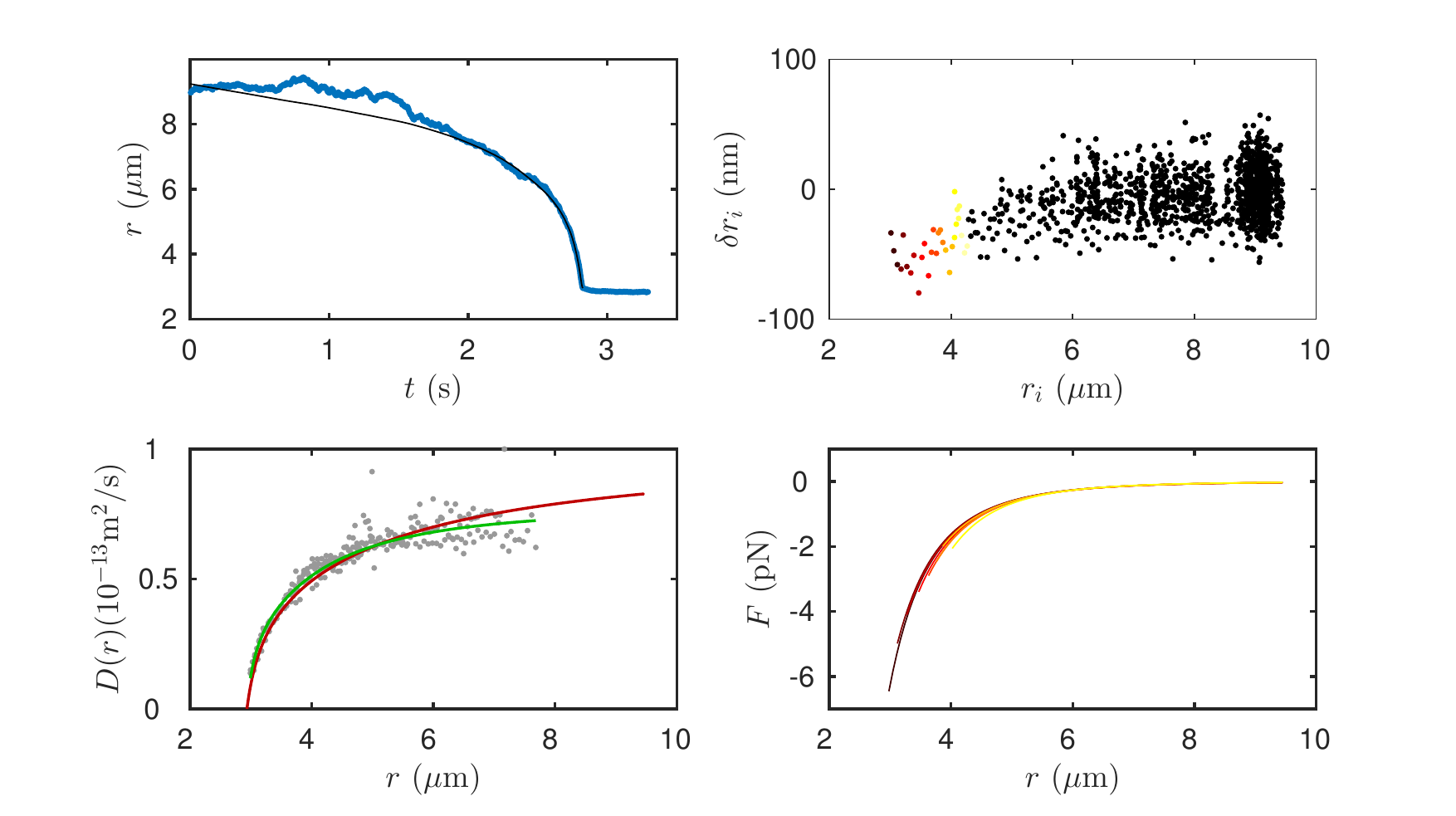}
\centering
\setlength{\unitlength}{1cm}
\put(-15.8,6.5)
{\includegraphics[height=1.3cm]{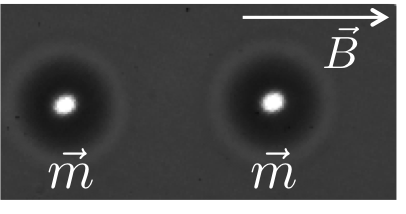}}
\put(-11,9)
{\large \text{(a)}}
\put(-7.7,9)
{\large \text{(b)}}
\put(-15.7,4)
{\large \text{(c)}}
\put(-7.7,4)
{\large \text{(d)}}
\vspace{-0.3cm}
\caption{\label{fig:trajcut} 
Experimental test of MLA  for spatially varying force and diffusion coefficient.
(a) Trajectory of the separation between two paramagnetic beads in a magnetic field (Inset).
(b) Corresponding displacements as a function of separation for the part of the trajectory before contact.
(c) Grey dots: diffusion coefficients obtained from the variance of the displacements ($3 \times 10^5$ time points).
Red curve: diffusion profile obtained from MLA of the trajectory shown in (b), (1400 time points).
Green curve: computational solution for two particles near a wall.
(d) Force profiles obtained from MLA of the trajectory shown in (b) with some of the colored points removed.
The color and spatial extent of the curve indicate the range of separations considered.
For example, the orange curve is the force fit to the trajectory including the black, light yellow and yellow $(r_i,\delta r_i)$ points, but excluding the dark orange, red and dark red points in (b).
}
\end{figure*}
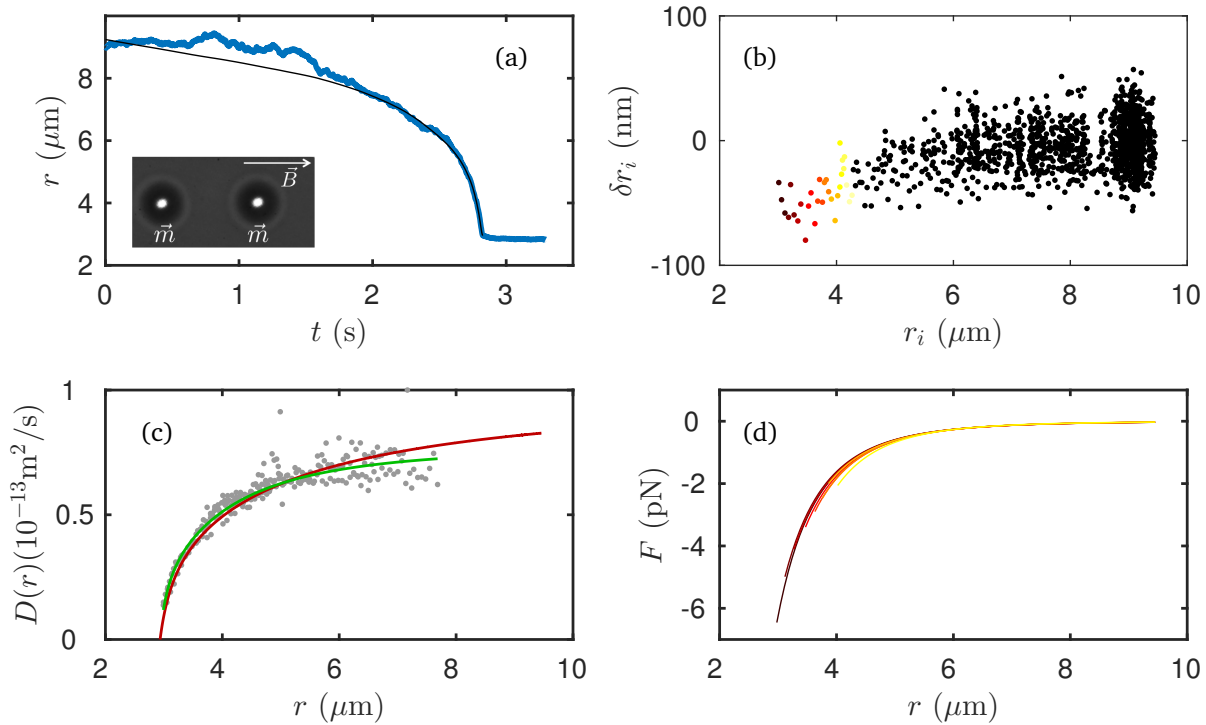

We apply MLA to a pair trajectory using functional forms described in eqn~(\ref{F_pl}) and eqn~(\ref{D_wall}), hence $\pmb{\alpha} = \lbrace \phi,\beta,D_0,R_0,D_h \rbrace$. The raw trajectory and displacement profile are shown in Fig.~\ref{fig:trajcut}a,b.  
The fitted trajectory, diffusion profile and force profile are shown in Fig.~\ref{fig:trajcut}a,c,d.

We investigate the reliability of the fits in two ways.
First, we compare the diffusion profile obtained from this single trajectory to an independent measurement.
Second, we look at how the MLA fit is modified when a few points of the trajectory are artificially removed: a robust fit ought to be resilient to small changes in the trajectory.

\subsubsection{Diffusion profile.}

To independently measure the diffusion profile, we tracked the displacements of the same particles in the absence of a magnetic field.
Here, there is no long range conservative interaction between the particles.
We record the positions of the particles at over $3\times 10^5$ timepoints, which allows us to thoroughly sample the variance of the displacements at randomly-sampled separations from near contact to a few radii \cite{biancaniello}.
The diffusion profile obtained from the binning of this large dataset is presented in Fig.~\ref{fig:trajcut}c as the grey dots.
In addition, we calculated the height- and separation-dependent mutual diffusion coefficient with an exact numerical approach\cite{jerzy2005,jerzy2005a}.
By fitting this calculation to the data, we determined the height above the wall to be 65~nm, and obtained the profile shown as the green curve.
The experimental and theoretical diffusion profiles are in good agreement with the MLA estimate of the diffusion profile measured for the strongly-interacting particles in a magnetic field, shown in red.
Notably, the MLA estimate required only 1400 timepoints, about 200 times less data than the direct measurement of the displacement variance.
In the presence of the magnetic field, the trajectory length was limited by the strong attraction between the beads and there was not enough data to accurately sample the variance in each spatial bin.

\subsubsection{Fit resilience to trajectory sampling.}

We assess the robustness of MLA by investigating how the estimated force profile is modified when a few points are removed from the trajectory.

The displacements $\delta r_i$ as a function of separation $r$ for the part of the trajectory before contact (Fig.~\ref{fig:trajcut}a, $t \leq 2.8$~s) are presented in Fig.~\ref{fig:trajcut}b. 
The large displacements at small separations (colored points in Fig.~\ref{fig:trajcut}b) may be expected to dominate the force profile.  
To test this hypothesis, we manually removed some of these $(r_i,\delta r_i)$ points, and performed MLA of the truncated trajectories.
The resulting force profiles are compared in Fig.~\ref{fig:trajcut}d, where the full trajectory is dark red and the line colors approach yellow as the trajectory is more strongly truncated (the colors correspond to the last points included in Fig.~\ref{fig:trajcut}b).
Truncation of the data at small separations  does change the estimated force profile, but the effects are very small (of the order of a few percents), even for the shortest trajectory (light yellow), where most of data points close to contact have been removed.
This shows that MLA provides a robust fit, in the sense that small changes in a trajectory have only small effects on the results from the fit.

\section{Practical considerations}

In this section, we present some general guidelines regarding the practical implementation of MLA.

\subsection{Data acquisition \label{sec:data}}
Experimental trajectories of Brownian particles are typically acquired using video microscopy, which involves two key temporal parameters: the frame-to-frame time interval $\Delta t$, and the exposure time $\tau_{ex}$.
Finite exposure times are well known to cause systematic errors in the observed variance of displacements at short time intervals \cite{wong,savin}.
To mimic realistic experimental conditions, we undersample and blur the simulated trajectories of Section~\ref{sec:pairsim}.
We use a readily accessible time interval $\Delta t = 4$~ms and an exposure time $\tau_{ex}=1$~ms.
(To emulate finite exposure times, we simply average the simulated positions over a 1~ms interval.)
The results of MLA  estimates of the parameters from these trajectories are shown as blue histograms in Fig.~\ref{fig:simul}c-f.
They agree reasonably well with the input parameters, with small but significant systematic errors.  
Similarly, blurring causes small but significant systematic errors in the displacement and force profiles, shown by the blue bands in  Fig.~\ref{fig:simul}g,h.
The displacements tend to be more dispersed near contact, and the forces are systematically overestimated.  
To minimize these systematic errors due to finite camera exposure time, it is important that $\tau_{ex}$ be significantly smaller than $\Delta t$.

\subsection{Statistical error estimates}
\begin{figure}[h]
\includegraphics[]{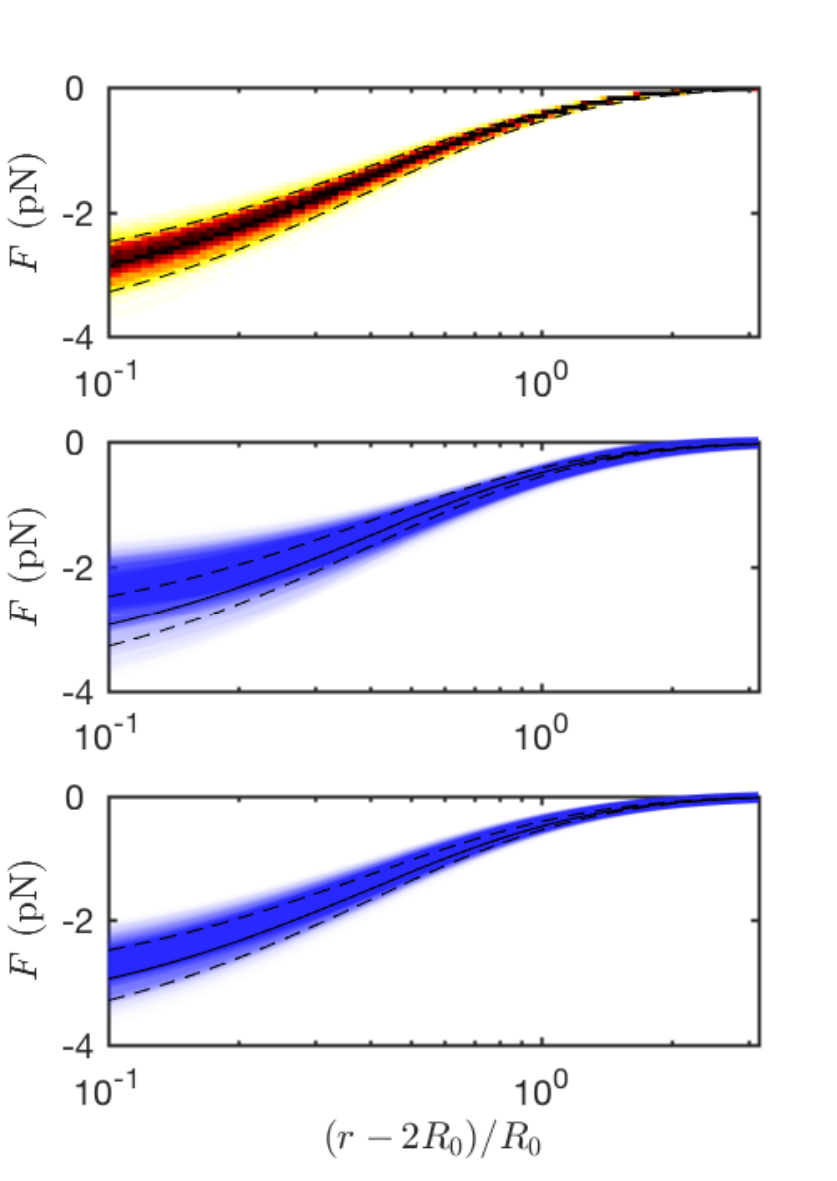}
\centering
\setlength{\unitlength}{1cm}
\put(-7,10.6)
{\large \text{(a)}}
\put(-7,7)
{\large \text{(b)}}
\put(-7,3.4)
{\large \text{(c)}}
\caption{\label{fig:lim} 
Statistical and systematic errors. 
Density of force curves obtained from MLA of simulated trajectories ($\Delta t = 0.1$~ms, $\tau_{ex} = 0$). 
Darker colors indicate higher density.
Solid black curve indicates the force or diffusion profile.
The two black dashed curves indicate the 90\% confidence interval, inferred using correct functional forms, from the density of curves at each separation.
(a) Histogram of the density of force profiles obtained from MLA of simulated trajectories using the correct functional forms for force and diffusion.
(b) Force profiles from MLA of simulated trajectories using the approximate force functional form $F^\star(r)$ described in Section~\ref{wrong}.
(c) Force profiles from MLA of simulated trajectories using the approximate diffusion functional form $D^\star(r)$ described in Section~\ref{wrong}.
}
\end{figure}

Statistical errors for a force profile obtained by MLA can be evaluated by simulating trajectories.
Consider an experimental trajectory fitted using functional forms $F_e(r)$ and $D_e(r)$, where MLA has returned $\pmb{\alpha_0}$ as the most likely estimate of the parameters.
One can then perform a Brownian dynamics simulation to simulate a large number of trajectories with the same time interval, length and spatial domains as the real experiments, using $F_e(r)$, $D_e(r)$, and $\pmb{\alpha_0}$. 
Each simulated trajectory can be analyzed to provide estimates of the parameters as well as the force and diffusion profiles.
The spreads in these values and fits capture the statistical uncertainty on the MLA estimates.  
To illustrate, we plot in Fig.~\ref{fig:lim}a the local density of force profile curves corresponding to the 1000 simulations described in Section \ref{sec:pairsim}.
From this two-dimensional histogram, we can calculate the 5th and 95th percentiles at each separation $r$ (black dashed lines), and hence recover a 90\% confidence interval.

\subsection{Functional form choice and resilience \label{wrong}}

When the force and diffusion profiles vary with position, MLA requires functional forms for each. 
In some cases, these functional forms may be known exactly from the literature, but in other cases, they are not.
In the latter case, one must use approximate functional forms $F^\star(r)$ and $D^\star(r)$.
Reasonable ones can be chosen based on some basic knowledge of the underlying physics, such as continuity, monotony, and limiting behavior.
To characterize the impact of using an incorrect but physically reasonable functional form, we analyzed trajectories from Section \ref{sec:pairsim} using functional forms different from the ones used for the simulations, eqn~(\ref{F_pl},\ref{D_isotr}).
MLA estimates of the force profile using an incorrect force profile 
$F^\star(r) = -\phi \exp(-r/\beta)$
are shown in Fig.~\ref{fig:lim}b.
Although they agree well with the input profile in the far-field, they significantly underestimate the force in the near-field.
Similarly, estimates of the force profile using an incorrect functional form for the diffusion profile 
$D^\star(r) = 2D_0 \left(1 - (3/2)r_0/r \right)$,
shown in Fig.~\ref{fig:lim}c, lead to small but significant errors in the force estimate.
However, in both cases we see that even with these wrong functional forms, the fitted force profiles remain essentially confined to the 90\% confidence interval, except maybe at very small separations with the exponential force profile, where the fitted profiles are on average $\sim 20$\% below what they should be (Fig.~\ref{fig:lim}b).

\section{Conclusions}

Maximum likelihood analysis is a powerful method to extract forces and diffusion coefficients from individual Brownian trajectories. 
It is fast and easy to implement.
It is particularly useful when the amount of data is limited and the force and diffusion profiles vary over space.
MLA is very efficient: it can achieve small statistical uncertainties with a modest amount of data, by exploiting the smooth spatial dependence of the force and diffusion profiles.
MLA can be very accurate, when the correct forms for the spatial dependence are employed.

In this paper and our previous implementation of the method, we considered experimental trajectories of micron-sized colloidal particles. 
Fundamentally, this technique could also be employed with much smaller objects, such as fluorescently labeled proteins.
Advanced microscopic techniques now allow for measurements of molecule positions with spacial resolution around 10~nm and time resolution better than 1~ms \cite{smt}.
However, due to photobleaching and other experimental difficulties, the recorded trajectories are typically very short, hence making the study of molecular interactions very difficult.
We believe that the development of MLA to the domain of single molecule tracking could help understand long-range interactions between biological components, hence clarifying the molecular cell processes which work by an interplay of diffusion and interaction.

This work was supported by the National Science Foundation (CBET 12-36086).
We would like to thank Jason Merrill, Robert Style, and Larry Wilen for helpful discussions.



\balance


\bibliography{ml_arX} 

\providecommand*{\mcitethebibliography}{\thebibliography}
\csname @ifundefined\endcsname{endmcitethebibliography}
{\let\endmcitethebibliography\endthebibliography}{}
\begin{mcitethebibliography}{21}
\providecommand*{\natexlab}[1]{#1}
\providecommand*{\mciteSetBstSublistMode}[1]{}
\providecommand*{\mciteSetBstMaxWidthForm}[2]{}
\providecommand*{\mciteBstWouldAddEndPuncttrue}
  {\def\EndOfBibitem{\unskip.}}
\providecommand*{\mciteBstWouldAddEndPunctfalse}
  {\let\EndOfBibitem\relax}
\providecommand*{\mciteSetBstMidEndSepPunct}[3]{}
\providecommand*{\mciteSetBstSublistLabelBeginEnd}[3]{}
\providecommand*{\EndOfBibitem}{}
\mciteSetBstSublistMode{f}
\mciteSetBstMaxWidthForm{subitem}
{(\emph{\alph{mcitesubitemcount}})}
\mciteSetBstSublistLabelBeginEnd{\mcitemaxwidthsubitemform\space}
{\relax}{\relax}

\bibitem[Clark \emph{et~al.}(1970)Clark, Lunacek, and Benedek]{clark1970}
N.~A. Clark, J.~A. Lunacek and G.~B. Benedek, \emph{Am. J. Phys.}, 1970,
  \textbf{38}, 575\relax
\mciteBstWouldAddEndPuncttrue
\mciteSetBstMidEndSepPunct{\mcitedefaultmidpunct}
{\mcitedefaultendpunct}{\mcitedefaultseppunct}\relax
\EndOfBibitem
\bibitem[Berne and Pecora(2000)]{DLS}
B.~J. Berne and R.~Pecora, \emph{Dynamic Light Scattering}, Dover Publications,
  Inc., Mineola, New York, 2000\relax
\mciteBstWouldAddEndPuncttrue
\mciteSetBstMidEndSepPunct{\mcitedefaultmidpunct}
{\mcitedefaultendpunct}{\mcitedefaultseppunct}\relax
\EndOfBibitem
\bibitem[Mason \emph{et~al.}(1997)Mason, Ganesan, van Zanten, Wirtz, and
  Kuo]{mason}
T.~G. Mason, K.~Ganesan, J.~H. van Zanten, D.~Wirtz and S.~C. Kuo, \emph{Phys.
  Rev. Lett.}, 1997, \textbf{79}, 3282\relax
\mciteBstWouldAddEndPuncttrue
\mciteSetBstMidEndSepPunct{\mcitedefaultmidpunct}
{\mcitedefaultendpunct}{\mcitedefaultseppunct}\relax
\EndOfBibitem
\bibitem[Roichman \emph{et~al.}(2008)Roichman, Sun, Roichman, Amato-Grill, and
  Grier]{roichman2008}
Y.~Roichman, B.~Sun, Y.~Roichman, J.~Amato-Grill and D.~G. Grier, \emph{Phys.
  Rev. Lett.}, 2008, \textbf{100}, 013602\relax
\mciteBstWouldAddEndPuncttrue
\mciteSetBstMidEndSepPunct{\mcitedefaultmidpunct}
{\mcitedefaultendpunct}{\mcitedefaultseppunct}\relax
\EndOfBibitem
\bibitem[Crocker and Grier(1994)]{crocker1994}
J.~C. Crocker and D.~G. Grier, \emph{Phys. Rev. Lett.}, 1994, \textbf{73},
  352\relax
\mciteBstWouldAddEndPuncttrue
\mciteSetBstMidEndSepPunct{\mcitedefaultmidpunct}
{\mcitedefaultendpunct}{\mcitedefaultseppunct}\relax
\EndOfBibitem
\bibitem[Perrin(1916)]{perrin}
J.~Perrin, \emph{{A}toms}, D. Van Nostrand company, New York, 1916\relax
\mciteBstWouldAddEndPuncttrue
\mciteSetBstMidEndSepPunct{\mcitedefaultmidpunct}
{\mcitedefaultendpunct}{\mcitedefaultseppunct}\relax
\EndOfBibitem
\bibitem[Prieve and Frej(1990)]{prieve1990}
D.~C. Prieve and N.~A. Frej, \emph{Langmuir}, 1990, \textbf{6}, 396\relax
\mciteBstWouldAddEndPuncttrue
\mciteSetBstMidEndSepPunct{\mcitedefaultmidpunct}
{\mcitedefaultendpunct}{\mcitedefaultseppunct}\relax
\EndOfBibitem
\bibitem[Chandrasekhar(1943)]{chandrasekhar1943}
S.~Chandrasekhar, \emph{Rev. Mod. Phys.}, 1943, \textbf{15}, 1\relax
\mciteBstWouldAddEndPuncttrue
\mciteSetBstMidEndSepPunct{\mcitedefaultmidpunct}
{\mcitedefaultendpunct}{\mcitedefaultseppunct}\relax
\EndOfBibitem
\bibitem[Sainis \emph{et~al.}(2007)Sainis, Germain, and Dufresne]{sainis2007}
S.~K. Sainis, V.~Germain and E.~R. Dufresne, \emph{Phys. Rev. Lett.}, 2007,
  \textbf{99}, 018303\relax
\mciteBstWouldAddEndPuncttrue
\mciteSetBstMidEndSepPunct{\mcitedefaultmidpunct}
{\mcitedefaultendpunct}{\mcitedefaultseppunct}\relax
\EndOfBibitem
\bibitem[Merrill \emph{et~al.}(2010)Merrill, Sainis, B\l{}awzdziewicz, and
  Dufresne]{merrill2010}
J.~W. Merrill, S.~K. Sainis, J.~B\l{}awzdziewicz and E.~R. Dufresne, \emph{Soft
  Matter}, 2010, \textbf{6}, 2187\relax
\mciteBstWouldAddEndPuncttrue
\mciteSetBstMidEndSepPunct{\mcitedefaultmidpunct}
{\mcitedefaultendpunct}{\mcitedefaultseppunct}\relax
\EndOfBibitem
\bibitem[Evans \emph{et~al.}(2016)Evans, Hollingsworth, and Grier]{evans2016}
D.~J. Evans, A.~D. Hollingsworth and D.~G. Grier, \emph{Phys. Rev. E}, 2016,
  \textbf{93}, 042612\relax
\mciteBstWouldAddEndPuncttrue
\mciteSetBstMidEndSepPunct{\mcitedefaultmidpunct}
{\mcitedefaultendpunct}{\mcitedefaultseppunct}\relax
\EndOfBibitem
\bibitem[Sarfati and Dufresne(2016)]{sarfati}
R.~Sarfati and E.~R. Dufresne, \emph{Phys. Rev. E}, 2016, \textbf{94},
  012604\relax
\mciteBstWouldAddEndPuncttrue
\mciteSetBstMidEndSepPunct{\mcitedefaultmidpunct}
{\mcitedefaultendpunct}{\mcitedefaultseppunct}\relax
\EndOfBibitem
\bibitem[Stamou \emph{et~al.}(2000)Stamou, Duschl, and Johannsmann]{stamou}
D.~Stamou, C.~Duschl and D.~Johannsmann, \emph{Phys. Rev. E}, 2000,
  \textbf{62}, 5263\relax
\mciteBstWouldAddEndPuncttrue
\mciteSetBstMidEndSepPunct{\mcitedefaultmidpunct}
{\mcitedefaultendpunct}{\mcitedefaultseppunct}\relax
\EndOfBibitem
\bibitem[Biancaniello and Crocker(2006)]{biancaniello}
P.~L. Biancaniello and J.~C. Crocker, \emph{Rev. Sci. Instr.}, 2006,
  \textbf{77}, 113702\relax
\mciteBstWouldAddEndPuncttrue
\mciteSetBstMidEndSepPunct{\mcitedefaultmidpunct}
{\mcitedefaultendpunct}{\mcitedefaultseppunct}\relax
\EndOfBibitem
\bibitem[Du \emph{et~al.}(2014)Du, Toffoletto, and Biswal]{biswal}
D.~Du, F.~Toffoletto and S.~L. Biswal, \emph{Phys. Rev. E}, 2014, \textbf{89},
  043306\relax
\mciteBstWouldAddEndPuncttrue
\mciteSetBstMidEndSepPunct{\mcitedefaultmidpunct}
{\mcitedefaultendpunct}{\mcitedefaultseppunct}\relax
\EndOfBibitem
\bibitem[Dufresne \emph{et~al.}(2000)Dufresne, Squires, Brenner, and
  Grier]{dufresne2000}
E.~R. Dufresne, T.~M. Squires, M.~P. Brenner and D.~G. Grier, \emph{Phys. Rev.
  Lett.}, 2000, \textbf{85}, 3317\relax
\mciteBstWouldAddEndPuncttrue
\mciteSetBstMidEndSepPunct{\mcitedefaultmidpunct}
{\mcitedefaultendpunct}{\mcitedefaultseppunct}\relax
\EndOfBibitem
\bibitem[Bhattacharya \emph{et~al.}(2005)Bhattacharya, B\l{}awzdziewicz, and
  Wajnryb]{jerzy2005}
S.~Bhattacharya, J.~B\l{}awzdziewicz and E.~Wajnryb, \emph{Physica A}, 2005,
  \textbf{356}, 294\relax
\mciteBstWouldAddEndPuncttrue
\mciteSetBstMidEndSepPunct{\mcitedefaultmidpunct}
{\mcitedefaultendpunct}{\mcitedefaultseppunct}\relax
\EndOfBibitem
\bibitem[Bhattacharya \emph{et~al.}(2005)Bhattacharya, B\l{}awzdziewicz, and
  Wajnryb]{jerzy2005a}
S.~Bhattacharya, J.~B\l{}awzdziewicz and E.~Wajnryb, \emph{J. Fluid. Mech.},
  2005, \textbf{541}, 263\relax
\mciteBstWouldAddEndPuncttrue
\mciteSetBstMidEndSepPunct{\mcitedefaultmidpunct}
{\mcitedefaultendpunct}{\mcitedefaultseppunct}\relax
\EndOfBibitem
\bibitem[Wong and Halvorsen(2006)]{wong}
W.~P. Wong and K.~Halvorsen, \emph{Optics Express}, 2006, \textbf{14},
  12517\relax
\mciteBstWouldAddEndPuncttrue
\mciteSetBstMidEndSepPunct{\mcitedefaultmidpunct}
{\mcitedefaultendpunct}{\mcitedefaultseppunct}\relax
\EndOfBibitem
\bibitem[Savin and Doyle(2005)]{savin}
T.~Savin and P.~S. Doyle, \emph{Biophysical Journal}, 2005, \textbf{88},
  623\relax
\mciteBstWouldAddEndPuncttrue
\mciteSetBstMidEndSepPunct{\mcitedefaultmidpunct}
{\mcitedefaultendpunct}{\mcitedefaultseppunct}\relax
\EndOfBibitem
\bibitem[Kusumi \emph{et~al.}(2014)Kusumi, Tsunoyama, Hirosawa, Kasai, and
  Fujiwara]{smt}
A.~Kusumi, T.~A. Tsunoyama, K.~M. Hirosawa, R.~S. Kasai and T.~K. Fujiwara,
  \emph{Nature Chemical Biology}, 2014, \textbf{10}, 524--532\relax
\mciteBstWouldAddEndPuncttrue
\mciteSetBstMidEndSepPunct{\mcitedefaultmidpunct}
{\mcitedefaultendpunct}{\mcitedefaultseppunct}\relax
\EndOfBibitem
\end{mcitethebibliography}
\bibliographystyle{rsc} 

\end{document}